# Miniaturized Low-power Electro-optic Modulator Based on Silicon Integrated Nanophotonics and Organic Polymers

Xingyu Zhang*[a], Amir Hosseini[b], Jingdong Luo[c], Alex K.-Y. Jen[c], and Ray T. Chen*[a]

[a] University of Texas at Austin, 10100 Burnet Rd, MER 160, Austin, TX 78758, USA;
[b] Omega Optics, Inc., 8500 Shoal Creek Blvd, Austin, TX 78757, USA;
[c] University of Washington, 302 Roberts Hall, Seattle, Washington 98195, USA.

**ABSTRACT**

We design and demonstrate a compact, low-power, low-dispersion and broadband optical modulator based on electro-optic (EO) polymer refilled silicon slot photonic crystal waveguide (PCW). The EO polymer is engineered for large EO activity and near-infrared transparency. The half-wave switching-voltage is measured to be $V_\pi=0.97\pm0.02$V over optical spectrum range of 8nm, corresponding to a record-high effective in-device $r_{33}$ of 1190pm/V and $V_\pi\times L$ of $0.291\pm0.006$V$\times$mm in a push-pull configuration. Excluding the slow-light effect, we estimate the EO polymer is poled with an ultra-high efficiency of 89pm/V in the slot. In addition, to achieve high-speed modulation, silicon PCW is selectively doped to reduce RC time delay. The 3-dB RF bandwidth of the modulator is measured to be 11GHz, and a modulation response up to 40GHz is observed.

**Keywords:** modulator, electro-optic polymer, photonic crystal waveguide, slow light, silicon photonics, silicon organic hybrid technology, microwave photonics

## 1. INTRODUCTION

Electro-optic (EO) polymer based modulators in optical links are promising for low power consumption [1] and broad bandwidth operation [2]. The electro-optic coefficient ($r_{33}$) of EO polymers can be several times larger than that of lithium niobate. And also, polymers are spin-on films so they can be easily spincoated onto any substrate without size limit. In addition to conventional all-polymer devices [1, 2], combination of silicon photonics and EO polymer have shown to enable compact and high-performance hybrid integrated photonic devices [3], such as slot waveguide Mach-Zehnder Interferometer (MZI) modulators [4], slot waveguide ring-resonator modulators [5], and slot Photonic Crystal Waveguide (PCW) modulators [6-8]. The fabrication process of these devices involves the poling of the EO polymer at an elevated temperature. Unfortunately, the leakage current due to the charge injection through silicon/polymer interface significantly reduces the poling efficiency in narrow slot waveguides (slot width, $S_w<200$nm). Among the abovementioned structure, the slot PCW can support optical mode for $S_w$ as large as 320nm [9]. Such a wide slot was shown to reduce the leakage current by two orders of magnitude resulting in 5x improvement in the in-device $r_{33}$ compared to a slot PCW with $S_w=75$nm [9], while high optical confinement in the slot is still achieved with the help of efficient mode converters [10]. One problem remains among slot PCW modulators is their narrow operating optical bandwidth of <1nm [11-13] because of the high group velocity dispersion (GVD) in the slow-light optical spectrum range. To broaden the operating optical bandwidth of PCW modulators, lattice shifted PCWs can be employed, where the spatial shift of certain holes can modify the structure to provide low-dispersion slow light [14-18].

In this letter, we report a symmetric MZI modulator based on band-engineered slot PCW refilled with EO polymer, SEO125 from Soluxra, LLC. SEO125 exhibits exceptional combination of large EO activity, low optical loss, and good temporal stability. Its $r_{33}$ value of poled thin films is around 125pm/V at the wavelength of 1310 nm, which is measured by the Teng–Man reflection technique. The design and synthesis of SEO125 encompasses recent development of highly efficient nonlinear optical chromophores with a few key molecular and material parameters, including large β values, good near-infrared transparency, excellent chemical- and photo-stability, and improved processability in polymers [19]. Using a band-engineered EO polymer refilled slot PCW with $S_w$=320nm, we demonstrate a slow-light enhanced effective in-device $r_{33}$ of 1190pm/V over 8nm optical spectrum range. Excluding the slow-light effect, we estimate in-device material' $r_{33}$ of 89pm/V for SEO125 in the slot that show 51% improvement compared to the results (59pm/V) in [9]. In addition, benefiting from the reduced RC time delay via silicon doping [20], the measured 3-dB RF bandwidth of the modulator is 11GHz, and modulation response up to 40GHz is observed.

*xzhang@utexas.edu; phone 1 512-471-4349; fax 1 512 471-8575
*raychen@uts.cc.utexas.edu; phone 1 512-471-7035; fax 1 512 471-8575

## 2. DESIGN

Our optical modulator is a symmetric Mach-Zehnder interferometer (MZI), with slot PCWs incorporated in both the arms. A schematic of the device on silicon on insulator (SOI) (Si thickness=250nm, oxide thickness=3μm) is shown in Figs. 1 (a) and (b). The slot and holes of the PCWs are infiltrated with EO polymer with a refractive index, n=1.63 at 1550nm. The refractive index of the EO polymer can be changed by applying an electric field via Pockel's effect, and is given as $\Delta n=(1/2)r_{33}n^3V/S_w$, where $\Delta n$ is the change in refractive index of the EO polymer, $r_{33}$ is the EO coefficient, V is the applied voltage, $S_w$ is the slot width. The slot PCW is designed with a lattice constant, a=425nm, a hole diameter, d=300nm, slot width, $S_w$=320nm, and center-to-center distance between two rows adjacent to the slot, W=1.54(√3)a. In order to efficiently couple light from strip waveguides into the slot PCW, adiabatic strip-to-slot mode converters are designed, as shown in the inset of Fig. 1 (a) [21]. To address the issue of the narrow optical bandwidth of typical PCW modulators (<1nm at $n_g$>10) [11, 12], the lattices of the second and third rows of the PCW are longitudinally shifted with relative values of $S_2$ = -85nm, $S_3$ = 85nm [indicated by the arrows in Fig. 1 (b)]). As a result, a group index ($n_g$) of 20.4 (±10%) over about 8nm optical wavelength range is achieved, as shown by the black curve in Fig. 1 (c), enabling a relatively large optical bandwidth of the modulator. To make a smooth transition between group indices ($n_g$) of slot waveguides ($n_g$~3) and slot PCWs ($n_g$~20.4), group index tapers consisting of 16 periods of non-lattice-shifted PCW, as shown in Fig. 1 (b), are developed, in which W increases parabolically from W=1.45(√3)a to W=1.54(√3)a from the beginning to the end of the input group index taper [22], and the resulting group index transition is illustrated in Fig. 1 (c). Sub-wavelength grating (SWG) are designed to couple light into and out of the silicon strip [23], and multi-mode interference (MMI) couplers are used for beam splitting/combining [24].

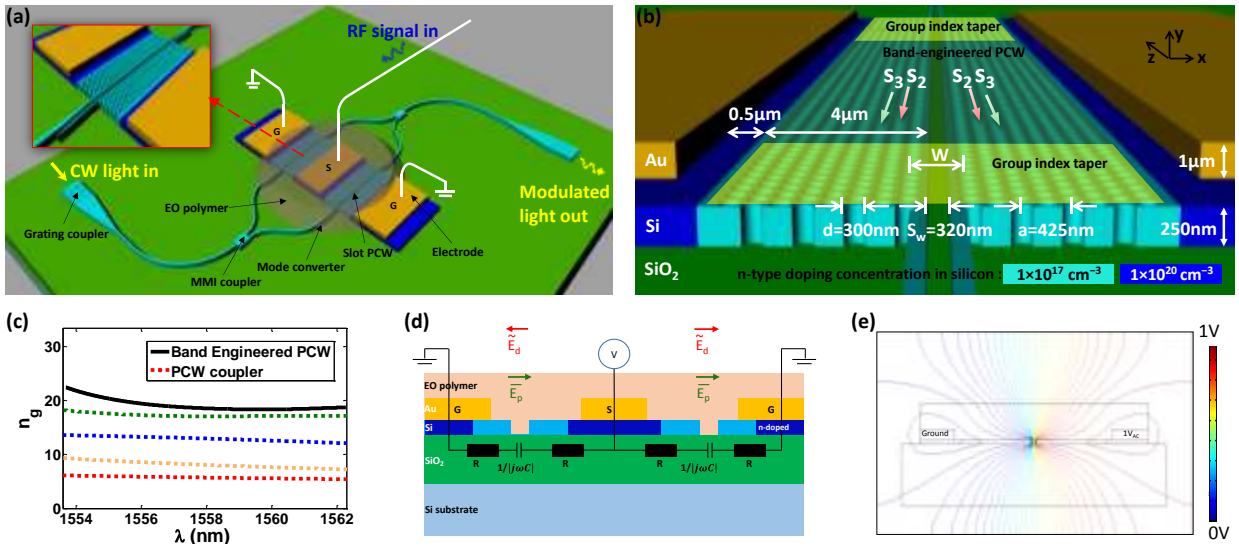

Fig. 1. (a) Three-dimensional schematic of the EO polymer infiltrated silicon slot PCW MZI modulator. The inset shows the magnified image of the silicon slot PCW on one arm of the MZI. (b) A tilted view of the slot PCW on one arm of the MZI, showing the device dimension, 2-level doping concentrations, group index taper region, and band-engineered PCW region. Note: the EO polymer is not shown here for better visualization. (c) The small variations of $n_g$ over about 8nm wavelength range, for the band-engineered slow-light PCW and PCW coupler. The lines of different colors represent the $n_g$ at different positions along the PCW coupler, indicating a smooth transition of $n_g$ from the beginning of the PCW taper to the band-engineered PCW. (d) Equivalent electrical circuit of the MZI modulator in a push-pull configuration. $E_d$: driving field, $E_p$: poling field. (e) Cross-sectional view of the simulated RF (10GHz) electric potential distribution across the doped silicon slot PCW filled with EO polymer.

Since our modulator is driven by lumped element electrodes, the main limiting factor for operation bandwidth is the RC time delay. To achieve broadband modulation, the silicon PCW is selectively implanted by n-type dopant (P+) with ion concentrations of $1\times10^{20}cm^{-3}$ and $1\times10^{17}cm^{-3}$ [8], as shown in Fig. 1 (b), so that the resistivity of silicon region is reduced to $9\times10^{-6}\Omega\cdot m$ and $9\times10^{-4}\Omega\cdot m$, respectively [25]. The relatively lower concentration of $1\times10^{17}cm^{-3}$ in the waveguide region is chosen to avoid significant impurity-induced scattering optical loss [26, 27]. Fig. 1 (d) shows a simplified equivalent circuit, in which the slot can be represented by a capacitor C and the silicon PCW region by R. Based on this equivalent circuit, as the modulation frequency increases, the percentage of electric potential dropped across the slot is supposed to decrease due to the reduced slot impedance. The low silicon resistivity can help increase the electric field inside the slot at high frequencies. The electric potential distribution at 10GHz in one arm is simulated by COMSOL Multiphysis as shown in Fig. 1 (e), and the simulation results show that over 90% of electric potential is dropped across the 320nm-wide slot at 10GHz. Both the optical field and modulation RF field are concentrated in the slot, enabling a large field interaction factor between them, and thus providing an efficient modulation at high modulation frequency. Based on simulations performed in Lumerical Device software, the total resistance of the 300μm-long silicon PCW is 189 Ohms, and the slot capacitance is as small as 39fF. Thus, the theoretical 3-dB modulation bandwidth of the MZI modulator is estimated to be $1/(2\pi RC)=22GHz$. The modulator is driven in a push-pull configuration as shown Figs. 1 (a) and (d).

## 3. FABRICATION

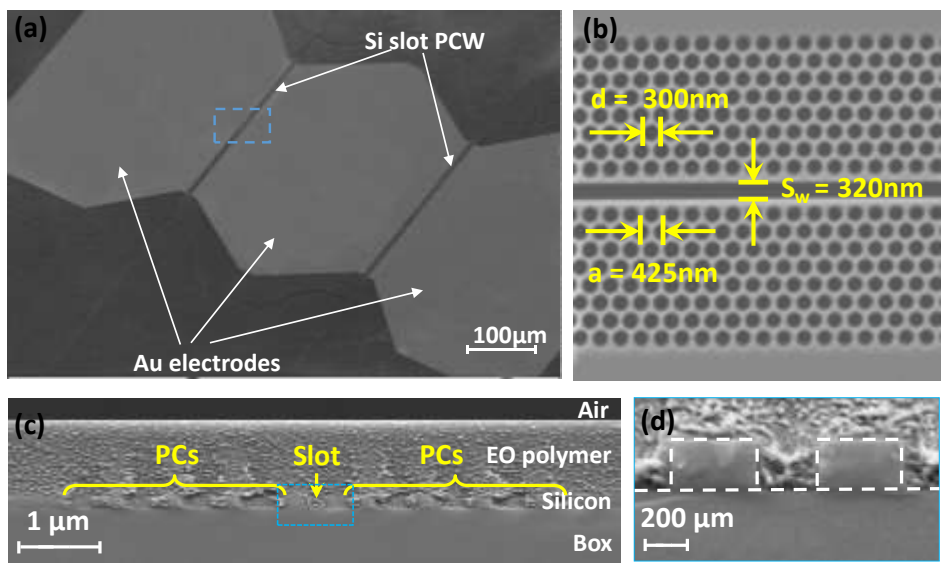

Fig. 2. SEM images of the fabricated device. (a) Tiled view of a local area of silicon slot PCW modulator. (b) Top view of slot PCW area. (c) Cross-sectional view of the EO polymer refilled silicon slot PCW. PCs: photonic crystals. (d) Zoom-in image of the dashed square area in (c).

The fabrication procedure starts with an SOI wafer with 250nm-thick top silicon. All the silicon photonic circuitries are fabricated using electron-beam lithography and reactive ion etching (RIE) in a single patterning/etching step. Then, the patterned silicon slot PCW is selectively doped by ion implantation, followed by rapid thermal annealing. Next, the 1μm-thick gold electrode with 5nm-thick chromium adhesion layer is deposited by using photolithography, electron-beam evaporation, and lift-off. Figs. 2 (a) and (b) show SEM images of the fabricated device. Next, the EO polymer, SEO125, is formulated and infiltrated into the slot PCW by spincoating. The silicon PCW regions including holes and the slot are fully covered by EO polymer, as shown in the SEM image in Figs. 2 (c) and (d). A microscope image of the fabricated MZI is shown in Fig. 3 (a). Next, to activate the EO effect of the polymer, the sample is poled by an electric field of 100V/μm in a push-pull configuration at the glass transition temperature ($T_g$=150 °C) of the EO polymer, so that the chromophore dipoles in the polymer are noncentrosymmetrically aligned in an uniform direction. The leakage current as well as the hot plate temperature is monitored and shown in Fig. 3 (b). It can be seen that the maximum leakage current remains below 0.659nA, corresponding to leakage current density of $8.79A/m^2$

[=0.659nA/(300um*250nm)]. For comparison, the typical leakage current density of the EO polymer is 1-10A/m² in a thin film configuration [7, 28]. This poling result is repeatable and shows that the 320nm-wide slot dramatically reduces the leakage current that is known to be detrimental to the poling efficiency [29].

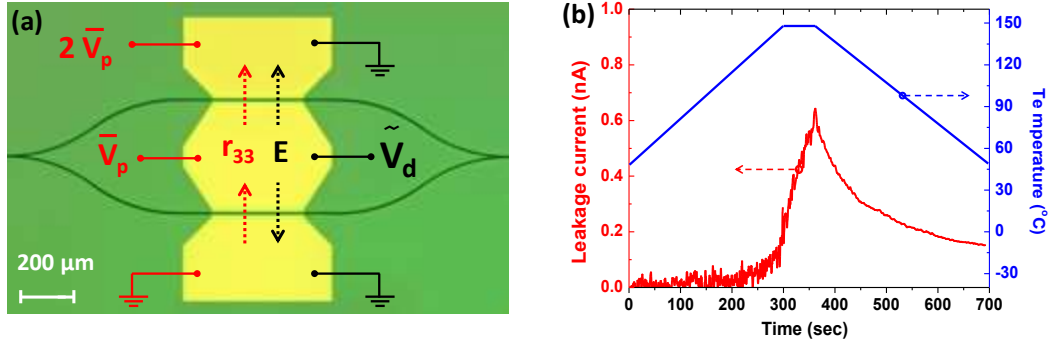

Fig. 3. (a) A microscope image of the top view of fabricated slot PCW MZI modulator. The red colored circuit connection indicates the push-pull poling configuration and induced $r_{33}$ direction, and the black colored circuit connection indicates the modulation configuration. $V_p$: poling voltage, $V_d$: diving voltage. (b) The temperature-dependent leakage current in the EO polymer poling process.

## 4. CHARACTERIZATION

Low-frequency modulation test is first implemented to measure the $V_\pi \times L$, which is a figure of merit (FOM) for optical modulators. TE-polarized light from a tunable laser source (1550nm) is coupled into and out of the device through grating couplers. The total optical insertion loss is 18dB. RF signals are applied to the electrodes as shown in Fig. 3 (a). The modulator is biased at the 3dB point and driven by a 100KHz triangular RF wave with a peak-to-peak voltage of 1.4V. The modulated output optical signal is sent to a photodetector and then displayed on a digital oscilloscope. The modulation frequency is within the bandwidth of the photodetector and the oscilloscope. From the output optical waveform measured by the digital oscilloscope, over-modulation is observed. The $V_\pi$ of the modulator is measured to be 0.973V from the transfer function of the over-modulated optical signal and the input RF signal on the oscilloscope, by finding the difference between the applied voltage at which the optical output is at a maximum and the voltage at which the optical output is at the following minimum [30]. The effective in-device $r_{33}$ is then calculated to be

$$r_{33-\text{effective}} = \frac{\lambda S_w}{n^3 V_\pi \sigma L} = 1190 \text{pm/V} \tag{1}$$

where, $\lambda$=1.55μm, $S_w$=320nm, n=1.63, L=300μm, σ=0.33 (confinement factor in the slot) calculated by simulation. This extraordinarily high $r_{33}$ value confirms the combined enhancing effects of slow light and an improved poling efficiency. This band-engineered 320nm slot PCW modulator also achieves very high modulation efficiency with $V_\pi \times L$=0.973V×300μm=0.292V×mm.

We also estimate the actual in-device $r_{33}$ excluding the slow-light effect using [14]

$$L = \frac{\lambda}{2\sigma n_g}\left(\frac{n}{\Delta n}\right) \tag{2}$$

where, $\Delta n = n^3 r_{33} V_\pi / (2 S_w)$. The estimated in-device $r_{33}$ is 89pm/V that is significantly larger than our previous work in [9] and is the highest poling efficiency demonstrated in a slot waveguide to the best of our knowledge. Considering the $r_{33}$ dispersion from the two-level model approximation [31], this value also represents nearly 100% poling efficiency that has been obtained in poled thin films of SEO125. Furthermore, for our lumped modulator without termination, the energy consumption is dominated by the capacitive load of the slot; therefore, the RF power consumption is estimated to be $P = 2\pi f C V_{rms}^2 \times 2 = 1.5$nW, where, f=100KHz (modulation frequency), C=0.01pF (slot capacitance) calculated by simulation, $V_{rms}=V_\pi/2/(\sqrt{2})$= 0.344V, and a factor of 2 is added due to the push-pull configuration. This ultralow power consumption benefits from the small capacitance due to large slot width and small $V_\pi$ due to slow-light enhanced high EO efficiency.

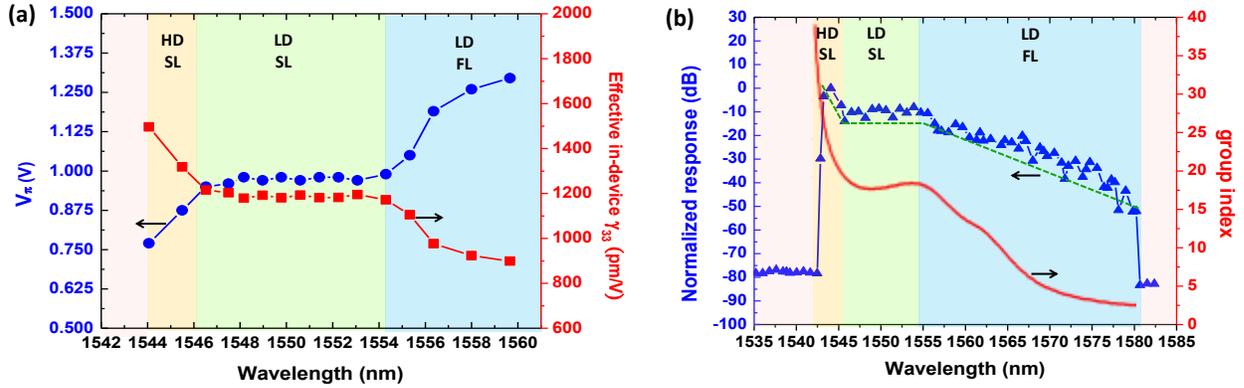

Fig. 4. (a) Measured $V_\pi$ and corresponding calculated effective in-device $r_{33}$ v.s. wavelength (at 100KHz). HD SL: high-dispersion slow-light; LD SL: low-dispersion slow-light; LD FL: low-dispersion fast-light. (b) Normalized device response v.s. wavelength (at 100KHz). The green dashed line indicates the trend of the response change over different wavelength. The simulated $n_g$ v.s. wavelength is also overlaid.

To demonstrate the wide optical spectrum range, the optical wavelength is tuned from 1544nm to 1560nm while all other testing conditions are fixed. The $V_\pi$ measured at different wavelength, as well as the corresponding calculated effective in-device $r_{33}$, is plotted in Fig. 4 (a). It can be seen that the $V_\pi$ is nearly constant, which is 0.97 ±0.02V, over optical spectrum range of 8nm (low-dispersion slow-light region: from 1546.5nm to 1554.5nm), corresponding to the effective in-device $r_{33}$ of 1190pm/V and $V_\pi \times L$ of 0.291 ± 0.006V×mm. We note that this $V_\pi \times L$ value is relative to a push-pull configuration. Relative to a single-arm modulator where the effective length of the MZI is the length of both arms together, $V_\pi \times (2L) = 0.582 \pm 0.012$V×mm is still a record low value. Furthermore, a small signal modulation test is done at $V_{pp}$<1V over a range of wavelength from 1535nm to 1582nm, while all other testing conditions remain the same. The modulated optical signal is converted to electrical signal by a photodetector and then measured by a microwave spectrum analyzer. The wavelength dependence of the normalized modulated optical signal is plotted in Fig. 4 (b). It can be seen that the defect-guided mode of slot PCW occurs from 1543nm to 1580nm. The maximum response occurs at the high-dispersion slow-light region (wavelength from 1543nm to 1546.5nm), because of the largest slow-light enhancement (largest $n_g$) in this region. The response is almost flat in the low-dispersion slow-light region (wavelength from 1546.5nm to 1554.5nm), because the slot PCW is band-engineered to have a nearly constant $n_g$ in this wavelength range. As the optical signal is tuned to longer wavelength (low-dispersion fast-light region: from 1554.5nm to1580nm), the device response becomes smaller due to decreasing $n_g$.

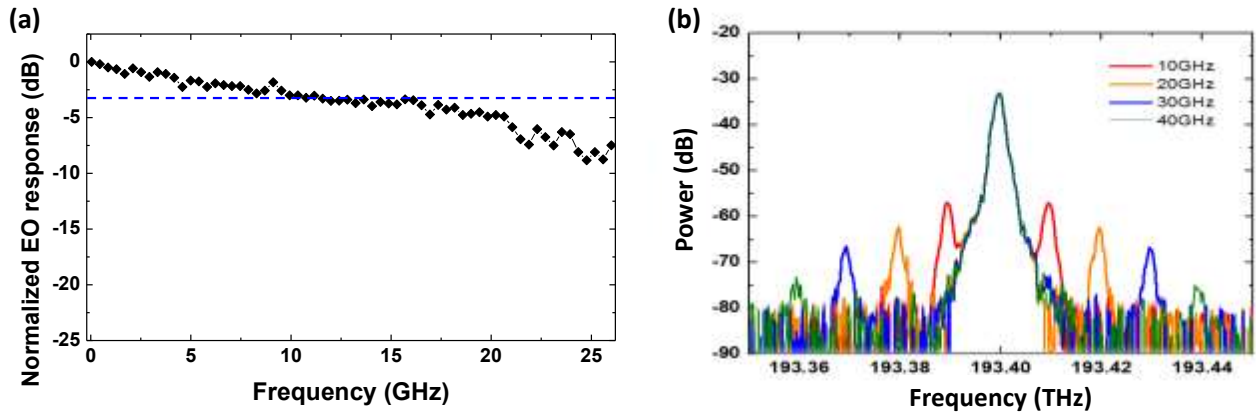

Fig. 5. (a) Measured normalized EO response of the modulator as a function of modulation frequency in a small-signal modulation test. The 3-dB bandwidth is measured to be 11GHz. (b) Measured optical transmission spectrum of the modulator operating at 10GHz, 20GHz, 30GHz and 40GHz.

Finally, the RF bandwidth is measured in a small signal modulation test. RF driving signal is provided by a network analyzer and applied onto the electrodes of the modulator via a ground-signal-ground (GSG) picoprobe. The modulated optical signal is amplified by an erbium doped fiber amplifiers (EDFA) and received by a high-speed photodetector, and then the detected electrical signal is measured using a microwave spectrum analyzer (MSA). The measured EO response of the device as a function of modulation frequency is shown in Fig. 5 (a), from which a 3-dB bandwidth of 11GHz is measured. Note that the upper frequency of this measurement is limited by the upper limit of our microwave spectrum analyzer (MSA), which is ~26GHz. Next, in order to overcome this measurement limit and measure the frequency response at frequencies over 26GHz, we perform another measurement using sideband detection technique [6, 32-34]. The optical output of the modulator is directly connected to the optical spectrum analyzer (OSA), and the optical transmission spectrum of the modulator is measured. When the modulator is driven by a high frequency RF signal, two sidebands appear, equally spaced around the main peak (193.4THz), in the transmission spectrum, as shown in Fig. 5 (b), where the horizontal axis is frequency and the spectral separation between the main peak and the sidebands indicates the modulation frequency. Fig. 5 (b) shows overlaid transmission spectra when the optical modulator is driven at 10GHz, 20GHz, 30GHz and 40GHz. At higher modulation frequencies, the power of the sidebands becomes lower due to the decreased electric potential drop across the slot, the reduced output power of the RF source, and the increased RF loss on the feeding cable and the probe. The observation of sideband signal at up to 40GHz indicates the achieved modulation at this high frequency.

## 5. CONCLUSION

In summary, we design, fabricate and characterize a band-engineered EO polymer refilled silicon slot PCW MZI modulator. The half-wave switching-voltage is measured to be $V_\pi=0.97\pm0.02$V over optical spectrum range of 8nm, corresponding to the slow-light enhanced effective in-device $r_{33}$ of 1190pm/V and $V_\pi\times L$ of $0.291\pm0.006$V×mm. Excluding the slow-light effect, we estimate the EO polymer is poled with a record-high EO activity of 89pm/V in the slot at the wavelength of 1550nm. By selectively doping silicon PCW, the RC time delay is effectively reduced, and hence high-speed modulation is achieved. Modulation response up to 40GHz is observed, with a measured 3-dB RF bandwidth of 11GHz. In our future work, the RF bandwidth can be further increased by applying a positive gate voltage from backside silicon substrate and generating accumulated electrons [35][36]. The optical loss of our modulator can be further reduced, such as by the design of low-loss PCW couplers [37] and improved coupling and packaging method [38]. The photochemical stability, a common issue for polymer based modulators, is expected to be improved by hermetically sealing of EO polymer in a robust packaging [39, 40]. Poled thin films of SEO125 have shown good temporal stability due to its relatively high $T_g$=150°C, and after the poling its EO coefficients were essentially unchanged under ambient conditions. While SEO125 is a newly developed material and its complete characterization in terms of performance and photo-stability is an ongoing effort. EO polymers with similar compositions have been demonstrated to have potential long-term stability by removing oxygen in the packaging of devices [41]. This modulator have potential applications ranging from optical interconnects [3] to microwave photonic sensing [42].

## ACKNOWLEDGEMENT

The authors would like to acknowledge the Air Force Research Laboratory (AFRL) for supporting this work under the Small Business Technology Transfer Research (STTR) program (Grant No. FA8650-12-M-5131) monitored by Dr. Robert Nelson and Dr. Charles Lee.